\documentclass[11pt]{article}

\usepackage{amsmath}
\usepackage{amssymb}
\usepackage{amsfonts}
\usepackage{latexsym}
\usepackage{amssymb}
\usepackage{amsfonts}
\usepackage{latexsym}

\newcommand{\eq}[1]{(\ref{#1})}

\newcommand{\n}[1]{\label{#1}}
\renewcommand{\baselinestretch}{1.1}
\def\beq{\begin{eqnarray}}
\def\eeq{\end{eqnarray}}
\def\ln{\,\mbox{ln}\,}

\def\Tr{\,\mbox{Tr}\,}

\def\al{\alpha}
\def\be{\beta}

\def\de{\delta}

\def\ep{\epsilon}

\def\la{\lambda}
\def\na{\nabla}

\def\si{\sigma}

\def\ph{\varphi}

\def\Ga{\Gamma}
\def\De{\Delta}

\def\Om{\Omega}

\begin{document}


\begin{center}

{\large{\bf Vacuum contribution of photons in the theory
with Lorentz and CPT-violating terms}}

\vskip 6mm


\textbf{Tib\'{e}rio de Paula Netto}$\,^{(a)}$
\footnote{
E-mail: tiberiop@fisica.ufjf.br}
\quad
\textbf{and}
\quad
\textbf{Ilya L. Shapiro}$\,^{(a,b)}$
\footnote{
E-mail: shapiro@fisica.ufjf.br}
\vskip 6mm

{\it (a)} \ \ Departamento de F\'{\i}sica, \ ICE, \
Universidade Federal de Juiz de Fora
\\ Juiz de Fora, \ 36036-330, \ MG, \ Brazil
\vskip 3mm

{\it (b)} \ \ Also at Tomsk State Pedagogical University and
at Tomsk State University, Tomsk, Russia
\end{center}

\vskip 12mm

\begin{quotation}
\noindent {\large {\it Abstract}}.
\quad
The photon contribution to the divergences and conformal
anomaly in the theory with Lorentz and CPT-violating terms
is evaluated. We calculate one-loop counterterms coming
from the integration over electromagnetic field and check
that they possess local conformal invariance. Furthermore,
conformal anomaly and anomaly-induced effective action are
calculated. It turns out that the new terms do not affect
the dynamics of conformal factor in the anomaly-driven
inflation (Starobinsky model) and its extensions. At the
same time, one can expect these terms to affect gravitational
wave equation and, in general, cosmic perturbations.
\vskip 3mm

{\it MSC:} \ 81T15, 
             83D05  
\vskip 3mm

{\it PACS:} \
          04.62.+v, 
          12.90.+b, 
          11.10.Gh, 
          04.50.Kd  

\end{quotation}
\vskip 4mm

\section{Introduction}

In the last decades there was a growing interest in the
theoretical and experimental aspects of the theories where Lorentz
and CPT symmetries are violated by special terms in the action of
quantum fields \cite{CPTL-reviews}. Many different tests have been
proposed in very different areas of physics, and there are good
perspectives to either discover such violation someday or benefit
from better understanding of physics which will result from the
continuously improving upper bounds on these new terms.
One of the badly explored yet aspects of the theories with Lorentz
and CPT-violating terms concerns cosmology\footnote{Except the
well-known work on baryogenesis, Ref. \cite{BCKP}.}. The present-day
state of art in this area is characterized by rapidly
growing precision, especially concerning the cosmic microwave
radiation (CMB), coming from the cosmic perturbations in the
early Universe. Therefore, it would be interesting to evaluate the
possibility of such violations, in particular at the inflationary
epoch. The early Universe can be seen as a subject of very
special interest, as far as Lorentz and CPT symmetries violation
is concerned. According to the formal Quantum Field Theory (QFT)
investigations \cite{beltor,guhesh,torsi}, torsion field, which is
one of the fields which may produce such violation, can not be
a propagating degree of freedom, because this would enter in
conflict with the unitarity of the theory at the quantum level.
At the same time, torsion can exist as a composite field which
results from some symmetry breaking in space \cite{torsi,Kost-grav}.
One can suppose that similar situation holds for other Lorentz
and CPT symmetries violating parameters, such that they result
from certain phase transition. Then the situation may be quite
different now and in the inflationary or post-inflationary
epochs, because some physical processes restoring the space-time
symmetries could occur since that time. For example, some of the
symmetry violations in the Early Universe could result in the
anisotropy in the CMB, which is apparently observed by Planck
\cite{Planck}.
Many of the Lorentz- and CPT-violating terms may lead to anisotropy
in the cosmological perturbations. Then, after these terms disappear
due to some kind of symmetry restoration, their imprint remains in
the CMB spectrum. Indeed, theoretical realization of this scheme
requires, first of all, a definition of the symmetry-breaking terms.

The natural next question is how to define the form of the
possible symmetries violation in the gravitational terms. One of
the possibilities is as follows. Assuming that the form of the
vacuum corrections should be derived from the quantum effects of
matter fields, it becomes obvious that the most relevant are the
contributions of photons, since all other particles are massive
and should decouple too early to produce a significant effect.
Therefore, the vacuum quantum contribution of photons is a natural
starting point for the formulation of possible CPT- and
Lorentz-violating terms in the gravitational sector. One more
comment is in order here. Apart from the quantum corrections,
one can introduce vacuum terms in CPT- and Lorentz-violating
theory in many different ways. For example, the general vacuum
action of gravity with torsion (small part of CPT- and
Lorentz-violating terms) includes 168 terms \cite{Christ-80}. Such
a great ambiguity makes it very difficult to expect any real
advances in this area. At the same time one can essentially
restrict the number of possible gravitational terms just by
introducing only those terms which can emerge as divergences
in the theory with Lorentz and/or CPT violating extensions in
the matter-fields sector.

The main purpose of the present work are contributions coming
from the massless photon field.
The derivation of one-loop divergences for massless conformal
invariant fields opens the way to the study of conformal anomaly
\cite{Duff94} and to the anomaly-induced effective
action of gravity \cite{rie,frts80}. The last is a useful,
compact analytic form of quantum correction, which can be
derived also in the presence of other fields, such as torsion
\cite{BuOdSh85,anhesh} and scalars \cite{Shocom,asta}. In this
work the anomaly-induced effective action will be extended to
the case of dimensionless
Lorentz and/or CPT violating parameter in the photon sector.
As an important example of cosmological application one can
consider the effect of the new terms to the anomaly-driven
inflation (Starobinsky model) \cite{star}. The complete
version of this model is based on the anomaly-induced
effective action of gravity, and can be extended to the cases
when other background fields are present
\cite{BuOdSh85,wave,book,Shocom,asta}.

The paper is organized as follows. Sect. 2 describes the technique
for deriving one-loop divergences in the electromagnetic theory
with the new external fields. Let us note that such a calculation
is not an easy thing to do, especially in the case of dimensionless
fields, as the reader will see in what follows. The method which
will be developed here
enables one to perform this and similar calculations up to the first
order in these fields, but, in principal, one can also go beyond this
order. Also in this section we briefly comment on the general structure
of renormalization in this theory. For a more extensive discussion of
this subject one can consult \cite{CPT-ren-2013}. Sect. 3 is
devoted to the technically difficult problem to prove the conformal
invariance of the bulky one-loop counterterms in the theory. After
this task is accomplished, the derivation of conformal anomaly
becomes a simple issue. Furthermore, in Sect. 4 we derive the
anomaly-induced effective action of gravity and also discuss
possible applications to inflation. Finally, in Sect. 5 we
draw our conclusions.

\section{Derivation of one-loop divergences}

Let us start with the action describing an extended version of
electromagnetic field with Lorentz and CPT symmetry breaking
terms. The corresponding action in flat space was formulated in
\cite{LCPT1}, and the minimal extension to the covariant form
is quite simple. The action is
\beq
\n{1}
S = \int d^4x \sqrt{-g}
\left\{ -\frac{1}{4} F_{\mu\nu} F^{\mu\nu}
-\frac{1}{4} k_F^{\mu\nu\al\be} \, F_{\mu\nu}
F_{\al\be} + \frac{1}{2}k_{AF}^\al \,
\epsilon _{\al \be \mu \nu} A^\be F^{\mu\nu} \right\}
\,,
\eeq
where $\,F_{\mu\nu} = 2\na_{[\mu} A_{\nu]}\,$
and parameters $k_F^{\mu\nu\al\be}$,
$k_{AF}^\al$ describe CPT and/or Lorentz violation.

For calculating the one-loop
divergences we shall apply the background
field method splitting (see, e.g., \cite{book} for
introduction),
\beq
\n{2}
A_\mu \rightarrow A_\mu + B_\mu
\,,
\eeq
where $B_\mu$ is the quantum field. The one-loop
effective action is given by the expression
\beq
\Ga^{(1)}_{div} = \frac{i}{2} \Tr \ln \hat{H}\big|_{div}
- i \Tr \ln \hat{H}_{gh}\big|_{div}
\,,
\eeq
here $\hat{H}$ is the operator of the
bilinear part of the action in quantum fields
and $\hat{H}_{gh}$ is the operator of the gauge (Faddeev-Popov)
ghosts term. Let us introduce the gauge-fixing term in the form
\beq
S_{gf} = \frac{1}{2\al} \int d^4x \sqrt{-g}
\, (\na_\mu B^\mu)^2
\,,
\eeq
with $\al$ is an arbitrary parameter of the gauge fixing. For this
choice of the gauge-fixing, the corresponding Faddeev-Popov ghosts
contribute only to the vacuum (metric dependent) sector of the
theory and these contributions do not depend on the new Lorentz
breaking parameters of the theory. We choose $\al=-1$ as a simplest
option for the practical calculations.

Replacing \eq{2} in the action \eq{1} one can find the bilinear
form of the action
\beq
S^{(2)} = \frac{1}{2} \int d^4 x \sqrt{-g}
\,\, B_\mu \, H^{\mu\nu} \, B_\nu
\,,
\eeq
where $\,H^{\mu\nu}=\hat{H}\,$ has the form
\beq
\hat{H} &=& \hat{H}_0 + \hat{H}_{AF} + \hat{H}_F \,,
\label{nm}
\\
\n{h0}
\hat{H}_0 &=& g^{\mu\nu}\Box - R^{\mu\nu}
\,,
\\
\hat{H}_{AF} &=& - 2 \, k_{AF}^\al  \,
\epsilon_\al ^{\,\,\,\,\,\mu\nu\be}
\na_\be - (\na_\be k_{AF}^\al )
\,\epsilon _\al ^{\,\,\,\,\,\mu\nu\be}
\n{AF}
\,,
\\
\hat{H}_{F} &=& - 2 \, k_F^{\mu(\al\be)\nu}
\, \na_\al \na_\be - 2 (\na_\al k_F^{\mu \al \be \nu})
\na_\be + k_F ^{ \mu \al \be \la}
R_{.\la \al\be}^{\nu}\,.
\eeq
The most important property of these formulas
is that operator (\ref{nm}) has a non-minimal
structure due to the term $\,k_F^{\mu(\al\be)\nu}\, \na_\al \na_\be$.
Then the standard Schwinger-DeWitt technique for deriving the
divergences can not be applied. Next, there is a well-elaborated
technique of dealing with non-minimal operators  \cite{bavi85},
but it works only in the cases when non-minimality can be
parameterized by some continuous parameter, such that one can
integrate over this parameter from zero (corresponding to the
minimal limit) and any given value. However, in the case of
(\ref{nm}) one meets a tensor field and not just a parameter.
Therefore, since this non-minimal
term in (\ref{AF}) has a non-standard form, the known technique
of dealing with non-minimal operators \cite{bavi85} can not be
applied too. We can conclude that the problem of our interest
lies beyond the limits of modern possibilities and hence its
complete solution is impossible.

In this situation one can try to consider certain approximation.
Let us assume that the parameters $\,k_F^{\mu\al\be\la}\,$ and
$\,k_{AF}^\al\,$ are small, such that the linear order in these
parameters will be sufficient for our purposes.
Indeed, the expansion can be taken to the next orders. In case
of the dimensional parameter $\,k_{AF}^\al\,$ such an expansion
will be finite, but for $\,k_F^{\mu\al\be\la}\,$ it can be
infinite. The general situation concerning renormalization
in the presence of parameters such as $\,k_{AF}^\al\,$ has
been recently described in \cite{CPT-ren-2013} and we will not
repeat it here completely, only give some necessary comments
at the end of this section. On the practical side we will
consider only linear order and, as the reader will observe,
it will be a technically difficult task.

So, for the sake of calculating the one-loop
divergences, let us first split the operator $\hat{H}$ into
minimal part $\hat{H}_m$ and the non-minimal part
$\hat{H}_{nm}$ and make the following transformation
\beq
\n{3}
\Tr \ln \hat{H}
&=&
\Tr \ln (\hat{H}_m + \hat{H}_{nm})
\,=\,
\Tr \ln \hat{H}_m
+ \Tr \ln (\hat{1} + \hat{H}^{-1}_{m} \hat{H}_{nm})
\\
\nonumber
&=&
\Tr \ln \hat{H}_m \,+\,
\Tr \hat{H}_{nm} \hat{H}^{-1}_0 \, +\, ...\,.
\eeq
In the last line we perform the expansion of
logarithm and take into account only terms in the
first order in the Lorentz and CPT violating parameters.
One can see that the first term in the last line of equation
\eq{3} can be directly calculated by the standard
Schwinger-DeWitt method \cite{dewitt}, while the second term
can be calculated by means of the universal functional traces
method (generalized Schwinger-DeWitt technique) of
Barvinsky and Vilkovisky \cite{bavi85}.

The minimal version of the operator \eq{nm} has been considered in
Ref. \cite{CPTLorentz10}, with the final result for the divergences
was obtained in the form
\beq
\n{trhm}
&&
\nonumber
\frac{i}{2} \Tr \ln \hat{H}_m\big|_{div}
- i \Tr \ln \hat{H}_{gh}\big|_{div} \,=
\\
\nonumber
&=&
- \,\frac{1}{\epsilon}\,
\int d^n x \,\, \mu^{n-4}\,\sqrt{-g} \Big\{
R_{\mu\nu} \na_\al \na_\be k_F^{\be \mu \al \nu}
- \frac{1}{6}\, R \, \na_\al \na_\be k_F^{\al \be}
\\
\nonumber
&+&
\frac{1}{3}\, R_{\mu\nu\al\be} \na^\be \na_\tau
k_F^{\tau \mu \al \nu}
- \frac{1}{12}\,k_{F}^{ \mu \nu \al \be } R R_{\mu \nu \al \be}
+ \frac{1}{2}\, k_{F}^{ \mu \al \be \tau}
R_{.\, \al \be \tau}^{\nu} R_{\mu\nu}
\Big\}
\\
&+&
\Ga^{(1)}_{vac}[g_{\mu\nu}]
\,.
\eeq
In the last formula we used a standard notation
$\,\ep =(4\pi)^2(n-4)\,$ for the parameter of
dimensional regularization and introduced a new notation
$\,k^{\mu \la \nu}_{F\,\,\,\,\,\la} \equiv k_F^{\mu\nu}$.
Also, $\,\Ga^{(1)}_{vac}[g_{\mu\nu}]\,$
is the divergent part of the metric-dependent
vacuum effective action of a massless vector field
(see, e.g., \cite{birdav,book}),
\beq
\Ga^{(1)}_{vac}[g_{\mu\nu}]  =
- \frac{1}{\epsilon }
\int d^n x \,\, \mu^{n-4}\, \sqrt{-g} \Big\{
\frac{1}{10}\, C^2 - \frac{31}{180}\, E
-\frac{1}{10}\, \Box R\Big\}
\,,
\eeq
with $C^2$ and $E$ representing the square
of the Weyl tensor and the Gauss-Bonnet topological
term (Euler density) respectively.

In the present work we shall go beyond the results of
\cite{CPTLorentz10}
and perform calculation for the case of the non-minimal operator.
Consider the contribution of the last term in the expression
\eq{3} for the divergences. For this calculation we need first
the inverse operator of \eq{h0}. As far as we are interested in
the divergences, the critically important observation is that,
from the viewpoint of power counting, the presence of the
dimensionless parameter $\,k_{F}^{\mu\nu\al\be\,}$ makes no changes.
Therefore, even in the presence of this parameter, the
counterterms will be given by the terms up to quadratic order
in curvature tensor, and it is safe to ignore higher order terms.
Then the inverse operator $\hat{H}_0^{-1}$ can be expressed as
\beq
\n{h0i}
\hat{H}_0^{-1} &=& (H_0^{-1})^\la_\nu =
\de^\la_\nu \frac{1}{\Box} + R^\la_\nu \frac{1}{\Box^2}
- 2 (\na^\rho R^\la_\nu) \na_\rho \frac{1}{\Box^3}
+ R^\la_\tau R^\tau_\nu \frac{1}{\Box^3} -
(\Box R^\la_\nu) \frac{1}{\Box^3}
\nonumber
\\
&+&
4 (\na^\rho \na^\si R^\la _\nu) \na_\rho \na_\si
\frac{1}{\Box^4} + {\cal O}(l^{-5})\,.
\eeq
In the last formula $\,{1/\Box}\,$ is the inverse
of d'Alembert operator and the last term $\,{\cal O}(l^{-5})\,$
indicates to an infinite series of omitted inessential terms of
higher background dimension $\,1/l$.

Using equation \eq{h0i} one can obtain the relation
\beq
\Tr \hat{H}_{nm} \hat{H}_0^{-1} &=&
- \,2 \Tr  k_F^{\mu (\al\be) \la}
\Big\{g_{\la\nu} \na_\al \na_\be
\frac{1}{\Box}
+ R_{\la \nu} \na_\al \na_\be \frac{1}{\Box^2}
+ (\na_\al \na_\be R_{\la \nu} ) \frac{1}{\Box^2}
\nonumber
\\
&+&
2 (\na_\al R_{\la \nu} ) \na_\be \frac{1}{\Box^2}
+ R_{\la \tau} R^\tau_\nu \na_\al \na_\be \frac{1}{\Box^3}
- 2 (\na^\rho  R_{\la \nu} ) \na_\al \na_\be
\na_\rho \frac{1}{\Box^3}
\nonumber
\\
&-&
4 (\na_\al \na^\rho  R_{\la \nu} )
\na_\be \na_\rho \frac{1}{\Box^3}
- (\Box R_{\la \nu}) \na_\al \na_\be \frac{1}{\Box^3}
\nonumber
\\
&+&
4 (\na^\rho \na^\si R_{\la \nu}) \na_\al \na_\be
\na_\rho \na_\si \frac{1}{\Box^4}
+ {\cal O} (R^3)\Big\} \,.
\n{8}
\eeq
The equation \eq{8} is already in the form
that allows us to apply the tables of universal functional
traces of generalized Schwinger-DeWitt technique \cite{bavi85}.
Using the functional traces formulas of this work,
each term of \eq{8} can be directly calculated.
As a result we obtain
\beq
\n{ti}
\Tr k_F^{\mu (\al\be) \la} (\na_\al \na_\be R_{\la \nu} )
\frac{1}{\Box^2}\Big|_{div} =
-\frac{2i}{\epsilon } \int d^nx \,\, \mu^{n-4}\, \sqrt{-g} \,
 k_F^{\mu \al\be \nu} \na_\al \na_\be R_{\mu \nu }
\,,
\eeq
\beq
\Tr k_F^{\mu (\al\be) \la} R_{\la \tau}
R^\tau_\nu \na_\al \na_\be \frac{1}{\Box^3}
\Big|_{div} =
\frac{i}{2\epsilon } \int d^nx \,\, \mu^{n-4}\, \sqrt{-g} \,
 k_F^{\mu\nu}
R_{\mu \la} R^\la_\nu
\,,
\eeq
\beq
- \Tr k_F^{\mu (\al\be) \la}(\Box R_{\la \nu})
\na_\al \na_\be \frac{1}{\Box^3}\Big|_{div}
=- \frac{i}{ 2 \epsilon } \int d^nx \,\, \mu^{n-4}\, \sqrt{-g} \,
k_F^{\mu\nu} \Box R_{\mu\nu}
\,,
\eeq
\beq
- 4 \Tr k_F^{\mu (\al\be) \la}
(\na_\al \na^\rho  R_{\la \nu} )
\na_\be \na_\rho \frac{1}{\Box^3}\Big|_{div}
=  \frac{2i}{\epsilon } \int d^nx \,\, \mu^{n-4}\, \sqrt{-g} \,
 k_F^{\mu \al \be \nu} \na_\al \na_\be
R_{\mu\nu}
\,,
\eeq
\beq
&&
4 \Tr k_F^{\mu (\al\be) \la}  (\na^\rho \na^\si
R_{\la \nu}) \na_\al \na_\be
\na_\rho \na_\si \frac{1}{\Box^4}
\Big|_{div}
\nonumber
\\
&=&
-\frac{2i}{\epsilon } \int d^nx \,\, \mu^{n-4}\, \sqrt{-g} \,
\Big\{\frac{1}{3} k_F^{\mu \al\be \nu}
\na_\al \na_\be R_{\mu \nu}
- \frac{1}{6} k_F^{\mu \nu} \Box R_{\mu \nu}\Big\}\,,
\eeq
\beq
2 \Tr k_F^{\mu (\al\be) \la} (\na_\al R_{\la \nu} )
\na_\be \frac{1}{\Box^2} \Big|_{div} = 0 \,,
\eeq
\beq
- 2 \Tr   k_F^{\mu (\al\be) \la} (\na^\rho
R_{\la \nu} ) \na_\al \na_\be \na_\rho
\frac{1}{\Box^3}
\Big|_{div} = 0
\,,
\eeq
\beq
&&
\Tr k_F^{\mu (\al\be) \la} R_{\la \nu}
\na_\al \na_\be \frac{1}{\Box^2}
\Big|_{div}
= - \frac{i}{\epsilon }
\int d^nx \,\, \mu^{n-4}\, \sqrt{-g} \,
\Big\{\frac{1}{3} k_F^{\mu\al\be\nu}
R_{\mu \nu} R_{\al \be}
\nonumber
\\
&&
+ \frac{1}{6} k_F^{\mu\nu} R
R_{\mu\nu} \Big \}
\,,
\eeq
\beq
\n{tf}
&&
\Tr k_F^{\mu (\al\be) \nu} \na_\al \na_\be
\frac{1}{\Box}\Big|_{div} =
- \frac{i}{\epsilon } \int d^nx \,\, \mu^{n-4}\, \sqrt{-g}
\Big\{ \,
\frac{1}{45} k_F^{\al \be} R^{\mu\nu} R_{\al \mu \be \nu}
+ \frac{19}{180} k_F^{\al \be} R_{\al \la \mu \nu}
R_\be^{\,\,\,\,\la\mu\nu}
\nonumber
\\
&&
- \frac{2}{45} k_F^{\al\be}
R_{\al \la} R^\la _\be + \frac{1}{18} k_F^{\al\be} R R_{\al \be}
+ \frac{1}{30} k_F^{\al \be} \Box R_{\al \be}
+ \frac{1}{10} k_F^{\al \be} \na_\al \na_\be R
\nonumber
\\
&&
+ \frac{1}{3} k_F^{\mu (\al\be) \nu}
R^\la_{\,.\mu\al\tau}R_{\la \nu \be .}^{ \quad \,\, \tau}
\,-\, k_F \, \Big(\frac{1}{180} R_{\mu\nu\al\be}^2 -
\frac{1}{180}R_{\mu\nu}^2 + \frac{1}{72}R^2
+ \frac{1}{30} \Box R \Big)
\Big\}
\,,
\eeq
where the notation $\,k_F \equiv g_{\mu\nu} k_{F}^{\mu\nu}$
has been introduced.
By using relations \eq{8}-\eq{tf}, one can obtain
\beq
\n{trhnm}
&&
\frac{i}{2} \Tr \ln \hat{H}_{nm} \hat{H}_0^{-1}\big|_{div} =
- \frac{1}{\epsilon } \int d^nx \, \mu^{n-4} \sqrt{-g}
\,\Big\{ \,
\frac{1}{45} k_F^{\al \be} R^{\mu\nu} R_{\al \mu \be \nu}
+ \frac{19}{180} k_F^{\al \be} R_{\al \la \mu \nu}
R_\be^{\,\,\,\,\la\mu\nu}
\\
&&
- \frac{49}{90} k_F^{\al\be}
R_{\al \la} R^\la _\be + \frac{2}{9} k_F^{\al\be} R R_{\al \be}
+ \frac{1}{5} R_{\al \be} \Box k_F^{\al \be}
+ \frac{1}{10} R \na_\al \na_\be k_F^{\al \be}
+ \frac{1}{3} k_F^{\mu (\al\be) \nu}
R^\la_{\,.\mu\al\tau}R_{\la \nu \be .}^{ \quad \,\, \tau}
\nonumber
\\
&&
+ \frac{2}{3}  R_{\mu \nu} \na_\al \na_\be
k_F^{\mu \al\be \nu}
+ \frac{1}{3} k_F^{\mu\al\be\nu}
R_{\mu \nu} R_{\al \be}
- k_F \, \Big(\frac{1}{180} R_{\mu\nu\al\be}^2
- \frac{1}{180}R_{\mu\nu}^2 + \frac{1}{72}R^2
+ \frac{1}{30} \Box R \Big)
\Big\}\,.
\nonumber
\eeq
Finally, from equations \eq{3}, \eq{trhm} and \eq{trhnm} we arrive
at the result for the one-loop divergences of effective action,
\beq
\n{div}
\Ga_{div}^{(1)} &=&
- \frac{1}{\epsilon } \int d^nx \,\, \mu^{n-4}\, \sqrt{-g}
\,K (g_{\mu\nu},k_F) \,\,+\,\, \Ga^{(1)}_{vac}[g_{\mu\nu}]\,,
\eeq
where
\beq
\n{K}
K (g_{\mu\nu},k_F) &=&
\frac{1}{45} k_F^{\al\be} R^{\mu\nu} R_{\al\mu\be\nu}
+ \frac{19}{180} k_F^{\al \be} R_{\al \la \mu \nu}
R_\be^{\,\,\,\,\la\mu\nu}
- \frac{49}{90} k_F^{\al\be}
R_{\al \la} R^\la _\be
+ \frac{2}{9} k_F^{\al\be} R R_{\al \be}
\nonumber
\\
&
+ &\frac{1}{5} R_{\al \be} \Box k_F^{\al \be}
- \frac{1}{15} R \na_\al \na_\be k_F^{\al \be}
+ \frac{1}{3} k_F^{\mu (\al\be) \nu}
R^\la_{\,.\mu\al\tau}R_{\la \nu \be .}^{ \quad \,\, \tau}
- \frac{1}{3}  R_{\mu \nu} \na_\al \na_\be k_F^{\mu \al\be \nu}
\nonumber
\\
&
+ &\frac{1}{3} k_F^{\mu\al\be\nu}
R_{\mu \nu} R_{\al \be}
+ \frac{1}{3} R_{\mu\nu\al\be} \na^\be \na_\la
k_F^{\mu \al \la \nu} - \frac{1}{12}
k_{F}^{ \mu \nu \al \be } R R_{\mu \nu \al \be}
\nonumber
\\
&
+& \frac{1}{2} k_{F}^{ \mu \al \be \la}
R_{. \al \be \la}^{\nu} R_{\mu\nu}
- k_F \, \Big(\frac{1}{180} R_{\mu\nu\al\be}^2
- \frac{1}{180}R_{\mu\nu}^2 + \frac{1}{72}R^2
+ \frac{1}{30} \Box R \Big)
\,.
\eeq
The expressions (\ref{div}), (\ref{K}) represent
the final result for the one-loop divergences in the linear
order in the parameter (field) $\,k_{F}^{\mu\nu\al\be}$.
Regardless of its bulky appearance, Eq. (\ref{K}) satisfies
some rigid constraints, as we shall see in the next section,
where (\ref{div}) will be used to calculate the conformal
anomaly.

As it was already said before, the result (\ref{div}), (\ref{K})
represents only the first term of an infinite series expansion in
the external field (space-dependent parameter)
$\,k_{F}^{\mu\nu\al\be}$. Since the classical term with
 $\,k_{F}^{\mu\nu\al\be}\,$ is not controlled by some fundamental
symmetry, at quantum level the situation here is not the same
as with external metric, which is also dimensionless, as
$\,k_{F}^{\mu\nu\al\be}\,$ is. However, in the metric case one
 can use general covariance and organize an infinite set of
 metric-dependent counterterms into a small amount of covariant
 expressions, namely in the $\,R^2_{\mu\nu\al\be}$, $\,R^2_{\mu\nu}$,
$\,R^2\,$ and $\,\Box R$ - terms (see, e.g., \cite{book,PoImpo} and
more formal recent discussion in \cite{Lavrov-CurRen}). In the
present case the situation is absolutely different, because $\,k_{F}^{\mu\nu\al\be}\,$ is the parameter of a purely
phenomenological origin and there is no fundamental symmetry
behind them. Therefore it is impossible to restore a full set
of counterterm from the lower-order expression such as (\ref{div}),
(\ref{K}) and, in case of a real interest, the next order terms
should be really calculated in an independent way. At the same
time, there are two pieces of exact information about higher order
terms. First, it is certain that these terms will have exactly
four derivatives, that means they will be quadratic in
curvature tensor components or have the structures like
$\na R \cdot k_F ... k_F \na k_F $,
or  $\na \na R \cdot k_F ... k_F$, or
$R \cdot k_F ... k_F \na\na k_F$, or
$\na R \cdot k_F ... \na k_F \na k_F $, or
$k_F ... k_F (\na k_F)^4$, or
$k_F ... k_F (\na k_F)^2\,(\na k_F)^2$, etc
(where we omitted all indices, of course). This feature is due
to the power counting-based arguments, which we already mentioned
before (see also \cite{CPT-ren-2013}).
The second certain property concerns the local conformal symmetry,
which will be
checked for (\ref{div}), (\ref{K}) in the next section. A standard
general argument shows that this symmetry will hold in all orders
in $\,k_{F}^{\mu\nu\al\be}$, and can be used for both verification
of quantum calculations and further applications.

The last observation is that, due to the complex calculations,
we did not derive the total derivative terms in $\Ga_{div}^{(1)}$.
This means, from the viewpoint of conformal anomaly, that we will
not be able to calculate the local terms of the anomaly-induced
effective action \cite{anomaly-AGS} and will take care only
about the (most relevant, usually) non-local part.

\section{Local conformal invariance and conformal anomaly}

The classical action of electromagnetic field in curved space
possesses local conformal invariance. This property is very
important, in particular it defined the equation of state
$\,P_r=\rho_r/3\,$ for  the radiation. The breaking of this
equation of state occurs only at quantum level due to the
conformal anomaly, and leads to a deformed equation of state
for radiation \cite{Dolgov93,radiana}. It is very important that
the classical action of electromagnetic field with Lorentz and
CPT symmetry breaking terms \eq{1} also possesses local
conformal invariance. In the present case this means that
the action of the theory does not change under the following
simultaneous transformation of the metric, of the vector
$A_{\mu}$ and of the parameter $\,k_F^{\mu\nu\al\be}$,
\beq
\n{tc}
g_{\mu\nu} \rightarrow g'_{\mu\nu} = g_{\mu\nu}\,e^{2\si}
\,,\quad
A_\mu \rightarrow A'_\mu = A_\mu
\,, \quad
k_F^{\mu\nu\al\be} \rightarrow
k_F^{'\mu\nu\al\be} = k_F^{\mu\nu\al\be}\, e^{-4\si}
\,,
\eeq
where $\si = \si (x)$. The local conformal invariance
of the action \eq{1} implies the vanishing trace of
energy-momentum tensor $T^\mu_\mu = 0$ in the  on-shell
limit. The same is true for the vacuum terms, if we do not
put there unnecessary non-conformal terms. However, the
situation changes dramatically, if we take quantum effects
onto account. At quantum level the classical action of vacuum
has to be replaced by the renormalized effective action
$\,\Ga_R$. Due the renormalization procedure the expectation
value of the trace $\langle T_\mu^\mu \rangle$ differs from
zero, which is called conformal (trace) anomaly \cite{Duff94}.

The renormalized one-loop effective action has the form
\beq
\Ga = S + \Ga^{(1)} + \De S
\,,
\eeq
where $\Ga^{(1)} = \Ga^{(1)}_{div} + \Ga^{(1)}_{fin}$
is  a direct quantum correction to the classical action
and $\De S$ is a local counterterm which is
called to cancel the divergent part of $\Ga^{(1)}$.
$\De S$ is the only source of the noninvariance
of the effective action, because classical action and
direct quantum contribution are conformal invariant.
Then the anomalous trace is
\beq
\n{con-an}
\langle T_\mu^\mu \rangle =
- \frac{2}{\sqrt{-g}} g_{\mu\nu}
\frac{\de \Ga_R}{\de g_{\mu\nu}}
\Bigg|_{n=4} =
- \frac{2}{\sqrt{-g}} g_{\mu\nu}
\frac{\de \De S}{\de g_{\mu\nu}}
\Bigg|_{n=4}
\,.
\eeq
The calculation of this expression can be done most simply
by using the conformal parametrization of the metric,
\beq
\n{metric-t}
g_{\mu\nu} = g'_{\mu\nu} \,e^{2\si} \,,
\eeq
where $g'_{\mu\nu}$ is the fiducial metric with fixed determinant
(this condition can be seen as purely technical and we can disregard
it after the derivation). One can easily prove the relation which
provides a simplest way to derive anomaly for new theories \cite{ConfPo},
\beq
\n{rel}
- \frac{2}{\sqrt{-g}} g_{\mu\nu}
\frac{\de A[g_{\mu\nu}]}{\de g_{\mu\nu}}
= - \frac{1}{\sqrt{-g'}} e^{-4\si}
\frac{\de A[g'_{\mu\nu} e^{2\si}]}{\de \si}
\Bigg|_{g'_{\mu\nu}\rightarrow g_{\mu\nu},
\si\rightarrow 0}
\,.
\eeq

In order to use these general results in our case, we need
first to prove that the conformal invariance of the new term,
\beq
\n{conI}
\sqrt{-g'} \,K (g'_{\mu\nu},k'_F)
&=& \sqrt{-g}\, K (g_{\mu\nu},k_F)\,,
\eeq
holds in the four dimensional space-time limit.
This is not a trivial task, from the technical side, so let
us present some details concerning the transformation rules.
For the one-parameter Lie group one can safely restrict the
consideration by the infinitesimal version of the
transformation \eq{tc}. Then, disregarding the higher orders
in $\si$ and superficial terms, after some long algebra we
arrive at the following transformation rules:
\beq
&&
(k_F^{\al\be}R^{\mu\nu}R_{\al\mu\be\nu})'
= (1-4\si)k_F^{\al\be}R^{\mu\nu}R_{\al\mu\be\nu}
+ 2 k_F^{\al \be} R_\be^\la \na_\al \na_\la \si
- k_F^{\al\be} R \na_\al \na_\be \si
\nonumber
\\
&&
- k_F R^{\al \be} \na_\al \na_\be \si
- k_F^{\al \be} R_{\al \be} \Box \si
+ 2 k_F^{\al\be}R_{\mu\al\be\nu} \na^\mu \na^\nu \si
+ \cdots
\eeq
\beq
&&
(k_F^{\al\be} R_{\al \la \mu \nu} R_{\be}^{\,.\,\la\mu\nu})'
= (1-4\si) k_F^{\al\be} R_{\al \la \mu \nu} R_{\be}^{\,.\,\la\mu\nu}
- 4 k_F^{\al \be} R_\be^\la  \na_\al \na_\la \si
\nonumber
\\
&&
+ 4 k_F^{\al\be} R_{\mu\al\be\nu} \na^\mu \na^\nu \si + \cdots
\eeq
\beq
(k_F^{\al\be}R_{\al \la}R^\la _\be)'
=(1-4\si) k_F^{\al\be}R_{\al \la}R^\la _\be
- 4 k_F^{\al \be} R_\be^\la \na_\al \na_\la \si
- 2 k_F^{\al\be} R_{\al \be} \Box \si
+ \cdots
\eeq
\beq
&&
(k_F^{\al \be} R_{\al \be} R )'
= (1-4\si)k_F^{\al \be} R_{\al \be} R
- 2 k_F^{\al \be} R \na_\al \na_\be \si
- 6 k_F^{\al \be} R_{\al\be} \Box \si
\nonumber
\\
&&
- k_F R \Box \si+ \cdots
\eeq
\beq
&&
(k_F^{\mu(\al\be)\nu}R^\la_{\,.\,\mu\al\tau}
R_{\la\nu\be\,.\,}^{\,\,\,\,\,\,\,\,\,\,\,\tau})'
= (1-4\si)k_F^{\mu(\al\be)\nu}R^\la_{\,.\,\mu\al\tau}
R_{\la\nu\be\,.\,}^{\,\,\,\,\,\,\,\,\,\,\,\tau}
+ k_F^{\al\be}R_{\mu\al\be\nu} \na^\mu \na^\nu \si
\nonumber
\\
&&
+ k_F^{\mu\al\be\nu}R_{\al\be} \na_\mu \na_\nu \si
- 6 k_F^{\mu\al\be\nu} R_{\la\mu\al\nu} \na^\la \na_\be \si
+ \cdots
\eeq
\beq
&&
(k_F^{\mu\al\be\nu} R_{\mu\nu} R_{\al\be})'
= (1-4\si)k_F^{\mu\al\be\nu} R_{\mu\nu} R_{\al\be} +
2 k_F^{\al \be} R_{\al \be} \Box \si
\nonumber
\\
&&
- 4 k_F^{\mu\al\be\nu} R_{\al \be} \na_\mu \na_\nu \si
+ \cdots
\eeq
\beq
&&
(k_F^{\mu\nu\al\be} R R_{\mu\nu\al\be})'
= (1-4\si)k_F^{\mu\nu\al\be} R R_{\mu\nu\al\be}
- 4 k_F^{\al \be} R \na_\al \na_\be \si
\nonumber
\\
&&
- 6 k_F^{\mu\nu\al\be}  R_{\mu\nu\al\be} \Box \si
+ \cdots
\eeq
\beq
&&
(k^{\mu\al\be\la} R^\nu_{\,.\,\al\be\la} R_{\mu\nu})'
= (1-4\si)k_F^{\mu\al\be\la} R^\nu_{\,.\,\al\be\la} R_{\mu\nu}
- 2 k_F^{\al \be} R_\be^\la \na_\al \na_\la \si
\nonumber
\\
&&
+ 2 k_F^{\mu\al\be\nu} R_{\al\be} \na_\mu \na_\nu \si
+ 2 k_F^{\mu\al\be\nu} R_{\la\mu\al\nu} \na^\la \na_\be \si
- k_F^{\mu\al\be\nu} R_{\mu\al\be\nu} \Box \si
+ \cdots
\eeq
\beq
&&
(R \na_\al \na_\be k_F^{\al \be})'
= (1-4\si) R \na_\al \na_\be k_F^{\al \be}
- 2 k_F^{\al\be} R \na_\al \na_\be \si
\nonumber
\\
&&
- 6 \na_\al \na_\be k_F^{\al \be} \Box \si
- 6 k_F^{\al\be} \na_\al R \,  \na_\be \si
+ k_F \na_\la R \,\na^\la  \si
+ \cdots
\eeq
\beq
&&
(R_{\al\be} \Box k_F^{\al\be})'
= (1-4\si)R_{\al\be} \Box k_F^{\al\be}
- 4 k_F^{\al \be} R_\be^\la \na_\al \na_\la \si
- 2 k_F^{\al \be} R_{\al\be} \Box \si
\nonumber
\\
&&
- 4  k_F^{\al\be} R_{\mu\al\be\nu} \na^\mu \na^\nu \si
- 2  \na_\al \na_\be k_F^{\al \be} \Box \si
- k_F \Box^2 \si
- 2 k_F^{\al \be} \na_\al R \,  \na_\be \si
\nonumber
\\
&&
- 2 k_F^{\al \be} \na_\la R_{\al\be} \na^\la  \si
+ 2 k_F^{\al \be} \na_\al R_{\la\be} \na^\la \si
- 2 k_F^{\al \be} \na^\tau R_{\tau\al\be\la} \na^\la \si
+ \cdots
\eeq
\beq
&&
( R_{\mu\nu} \na_\al \na_\be k_F^{\mu\al\be\nu} )'
= (1-4\si) R_{\mu\nu} \na_\al \na_\be k_F^{\mu\al\be\nu}
+ k_F^{\al \be} R_\be^\la \na_\al \na_\la \si
\nonumber
\\
&&
- k_F^{\mu\al\be\nu} R_{\al\be} \na_\mu \na_\nu \si
+ 2 k_F^{\mu\al\be\nu} R_{\la\mu\al\nu} \na^\la \na_\be \si
+ \na_\al \na_\be k_F^{\al \be} \Box \si
\nonumber
\\
&&
- k_F^{\al\be} \na_\la R_{\al\be} \na^\la \si
+ 2 k_F^{\al\be} \na_\al R_{\la\be} \na^\la \si
- 4 k_F^{\mu\al\be\nu} \na_\al R_{\mu\nu} \na_\be \si
\nonumber
\\
&&
- 2 k_F^{\mu\al\be\nu} \na_\be R_{\mu\la\al\nu} \na^\la \si
+ \cdots
\eeq
\beq
&&
(R_{\mu\nu\al\be} \na^\be \na_\la k_F^{\mu\al\la\nu})'
= (1-4\si) R_{\mu\nu\al\be} \na^\be \na_\la k_F^{\mu\al\la\nu}
+ k_F^{\al \be} R_\be^\la \na_\al \na_\la \si
\nonumber
\\
&&
- k_F^{\mu\al\be\nu} R_{\al\be} \si_{\mu\nu}
+ 2 k_F^{\mu\al\be\nu} R_{\la\mu\al\nu}  \na^\la \na_\be \si
+ \na_\al\na_\be k_F^{\al\be} \Box \si
+ k_F^{\al\be} \na_\al R_{\la\be} \na^\la \si
\nonumber
\\
&&
+ k_F^{\al \be} \na^\tau R_{\tau\al\be\la} \na^\la \si
- k_F^{\mu\al\be\nu} \na_\al R_{\mu\nu} \na_\be \si
+ 3 k_F^{\mu\al\be\nu} \na^\tau R_{\tau\mu\nu\al} \na_\be \si
\nonumber
\\
&&
- 2 k_F^{\mu\al\be\nu} \na_\be R_{\mu\la\al\nu} \na^\la \si
+ \cdots
\eeq
\beq
&&
\Big[k_F \, \Big(\frac{1}{180} R_{\mu\nu\al\be}^2
- \frac{1}{180}R_{\mu\nu}^2 + \frac{1}{72}R^2
+ \frac{1}{30} \Box R \Big)\Big]'
= (1- 4\si) k_F \, \Big(\frac{1}{180} R_{\mu\nu\al\be}^2
\nonumber
\\
&&
- \frac{1}{180}R_{\mu\nu}^2 + \frac{1}{72}R^2
+ \frac{1}{30} \Box R \Big)
- \frac{1}{45} k_F R^{\al\be} \na_\al \na_\be \si
- \frac{2}{9} k_F R \Box \si
- \frac{1}{5} k_F \Box^2 \si
\nonumber
\\
&&
- \frac{1}{15} k_F \na_\la R \, \na^\la \si
+ \cdots
\eeq
Substituting these formulas into \eq{K}, we
find the conformal invariance \eq{conI}.

By using Eqs. \eq{con-an}, \eq{rel} and \eq{conI},
one can easily find the conformal anomaly,
\beq
\n{ano}
\langle T_\mu^\mu \rangle
= -\frac{1}{(4\pi)^2}
\big[ w C^2 + b E + c \Box R + K(g_{\mu\nu},k_F) \big]\,,
\eeq
where the parameters  \ $w$, $b$, $c$ \ are, in our case,
\beq
w = \frac{1}{10}
\,, \quad
b = -\frac{31}{180}
\,, \quad
c = - \frac{1}{10}
\,.
\eeq

\section{Anomaly-induced effective action}

One can use the conformal anomaly \eq{ano} to construct an equation
for the finite part of the 1-loop correction to the effective action
\beq
\n{dif-e}
\frac{2}{\sqrt{-g}} g_{\mu\nu}
\frac{\de \Ga_{ind}}{\de g_{\mu\nu}}
= \frac{1}{(4\pi)^2}
\big[ w C^2 + b E + c \Box R + K(g_{\mu\nu},k_F) \big]\,.
\eeq
The solution of this equation is straightforward. The simplest
possibility is to parameterize metric as in \eq{metric-t}, separating
the conformal factor $\,\si(x)\,$ and rewrite the Eq. \eq{dif-e}
using \eq{rel}. The solution for the effective action is
\beq
\n{ano-action}
\Ga_{ind} &=& S_c[g_{\mu\nu}'] +
\frac{1}{(4\pi)^2} \int d^4 x \sqrt{-g'}
\Big\{ w \si C^2 + b \si \Big( E' - \frac{2}{3} \Box' R'
\Big) + 2 b \si \De'_4 \si
\nonumber
\\
&+&
\si K(g'_{\mu\nu},k'_F)
-\frac{3c+2b}{36}[R'-6(\na' \si)^2
- 6 \Box' \si]^2  \Big\}\,,
\eeq
where $\De_4$ is a fourth derivative conformal
covariant Paneitz operator, acting on dimensionless scalar
\beq
\De_4 = \Box^2 + 2 R^{\mu\nu} \na_\mu \na_\nu
- \frac{2}{3} R \Box + \frac{1}{3} (\na^\mu R) \na_\mu\,.
\eeq
$S_c[g_{\mu\nu}']=S_c[g_{\mu\nu}]\,$ in Eq. (\ref{ano-action})
is an arbitrary
conformal invariant functional of the metric, which serves as an
integration constant of Eq. (\ref{dif-e}). In the purely metric
theory this functional is irrelevant for the dynamics of the
conformal factor. Then, for the simplest cosmological applications,
the anomaly-induced expression can be seen as an exact effective
action. It is important that this term can be also ignored
when one is dealing with the black-hole applications
\cite{balsan,AMV} and gravitational waves \cite{star83,wave,HHR}.
In both cases the results obtained {\it without} this term
provide a very good fit with the ones obtained by other methods.
The reason for this output is that the rest of the action
(\ref{ano-action}) keeps full information about the UV limit
of the theory. In other works, it contains all the leading
logarithmic corrections, while for $\,S_c[g_{\mu\nu}]\,$
remain only sub-logarithmic parts.

When other background fields are present, the automatic
irrelevance of the term $\,S_c[g_{\mu\nu}]\,$ in the
zero-order cosmology does not hold,
because  $\,S_c[g_{\mu\nu}]\,$ may depend on these
fields, along with the metric. Our present situation belongs
to this class of theories \cite{anhesh,Shocom,asta}, because
this integration constant may depend also on
$\,k_F^{\al\be\mu\nu}$. This means
$\,S_c=S_c[g_{\mu\nu},\,k_F^{\al\be\mu\nu}]$. However, taking
into account the arguments presented above, we will not
really care about this term.

The expression \eq{ano-action} is the quantum correction to the
classical action. Let us note that the covariant forms of the
anomaly-induced action can be easily calculated on the basis of
Eq. (\ref{ano-action}), in both nonlocal  \cite{rie,frts80} and
local forms, the last uses auxiliary fields \cite{a,Mottola} (see
also \cite{ConfPo} for a review).

Let us give just a final result for the local form
of anomaly induced effective action, with the two auxiliary
scalar fields $\ph$ and $\psi$. Compared to the original formula
of \cite{a}, this expression has an extra term related to the
parameter $k_F^{\al\be\mu\nu}$,
\beq
\Ga_{ind} &=& S_c[g,\,k_f]
\,-\, \frac{3c+2b}{36(4\pi)^2}\,\int d^4 x \sqrt{g (x)}\,R^2(x)
\,+\,  \int d^4 x \sqrt{g (x)}\,\Big\{
\frac12 \,\ph\De_4\ph - \frac12 \,\psi\De_ 4\psi
\nonumber
\\
&+& \ph\,\left[\,\frac{\sqrt{-b}}{8\pi}\,\Big(E -\frac23\,{\Box}R\Big)\,
- \frac{1}{8\pi\sqrt{-b}}\,
\big(aC^2 + K(g_{\mu\nu},k_F)  \big)\,\right]
\cr
&+&
\frac{1}{8\pi \sqrt{-b}}\,\psi\,\big(aC^2
+  K(g_{\mu\nu},k_F)\big) \,\Big\}\,.
\label{finaction}
\eeq
The last form of the effective action is the most useful one
for dealing with Hawking radiation from black holes or exploring
the dynamics of gravitational waves on cosmological background.
In both cases one has to solve the equations for the auxiliary
fields $\ph$ and $\psi$ by implementing the appropriate
boundary conditions. After that it is possible to study the
energy-momentum tensor of vacuum in case of black holes
\cite{balsan,AMV} or explore the dynamics of gravitational
waves \cite{wave}. Indeed, for the homogeneous and isotropic
metrics there is no difference between the effective actions
(\ref{finaction}) and (\ref{ano-action}), they always give
the same dynamics of the conformal factor $\si$. Hence,
Eq. \eq{ano-action} is completely sufficient for exploring
the dynamics of the conformal factor, which we are going
to study in the rest of this section.

Consider possible applications of anomaly (\ref{ano}) and
the anomaly-induced effective action (\ref{ano-action}) to
inflation. The starting point should be the theory based
on the Einstein-Hilbert action with quantum correction
(\ref{ano-action}),
\beq
\n{totalaction}
S = - \frac{M_p^2}{16\pi} \int d^4x \sqrt{-g} \,R + \Ga_{ind}\,,
\eeq
where $M_p^2 = 1/G$ is the square of the Planck
mass and $\Ga_{ind}$ is the quantum correction \eq{ano-action}.
We look for an isotropic and homogeneous solution
\beq
g_{\mu\nu} &=&   a^2(\eta)\,g'_{\mu\nu}
\,,
\eeq
where $\eta$ is the conformal time
\beq
\n{metricprime}
ds'^2 = g'_{\mu\nu} dx^\mu dx^\nu
= d\eta^2 - \frac{dr^2}{1-k r^2} - r^2 d\Om
\eeq
and $k$ parameterizes the space-time curvature $\,k=0,\pm 1$.

The first observation concerning the effect of the parameter
$\,k_F^{\al\be\mu\nu}\,$ is its complete irrelevance for the
flat-space case $k=0$. The reason is that the effect of
Lorentz- and CPT-violating parameters is accumulated in the
scalar function $K=K (g'_{\mu\nu},k_F^{'\al\be\mu\nu})$. From
the definition of this function in (\ref{K}), it directly
follows that $\,K(\eta_{\mu\nu},k_F^{'\al\be\mu\nu}) = 0$.
Therefore, the only chance to observe some effect of the
Lorentz- and CPT-violating parameter $\,k_F^{\al\be\mu\nu}\,$
on the dynamics of conformal factor is related to the cases
$\,k=\pm 1$.

The direct calculation of the new term, induced by Lorentz-
and CPT-symmetry breaking term $\,K(g'_{\mu\nu},k'_F)$
requires some long algebra and we shall give only a final
result. It is relatively easy to show that all terms which
involve $\,g'_{o\nu}\,$ for $\,\nu=0,1,2,3\,$ give zero.
For the space indices $i,j=1,2,3$ one can show, by using
metric \eq{metricprime} in Eq. \eq{K}, the following
relation:
\beq
K (g'_{\mu\nu},k'_F)
&=&
k^2 k_F^{'ij} g'_{ij}
- \frac{1}{2} \,k^2 \, k_F^{'iklj}\, g'_{il}\, g'_{kj}
- \frac{1}{2}\, k^2\, k'_F\,.
\eeq
At this point we have to remember that the tensor
$\,k_F^{\al\be\mu\nu}\,$ has the same algebraic symmetries
as the Riemann tensor. \ According to the definitions, $\,k'_F=k_F^{'\mu\nu\al\be}g'_{\mu\al} g'_{\nu\be}\,$
and
$\,k_F^{'\nu\be}=k_F^{'\mu\nu\al\be}g'_{\mu\al}$,
it is not difficult to check that, finally,
$\,K (g'_{\mu\nu},k'_F)=0$. This means that the new term
with $\,K (g_{\mu\nu},k_F)\,$ gives no contribution to the
dynamics of the conformal factor in the theory
(\ref{totalaction}).

The negative result concerning the effect of the new terms
on the behavior of conformal factor of the metric does not
mean that there can not be other relevant effects. In particular,
one can expect the modifications of equations for cosmic
perturbations \cite{MuCh} and especially for the gravitational
waves. An important result concerning the dynamics of
traceless and transverse perturbations of the metric in
the theory (\ref{totalaction})  without the term
$\,K(g_{\mu\nu},k_F)$ is that there are no growing
modes in this theory \cite{star83,wave,HHR}. This fact
has important phenomenological consequences, including
relatively small role of tensor perturbations compared to
the scalar one (see, e.g., \cite{BrandMukh}). It would be
interesting to check whether the situation remains the same
or gets changed in the theory with by Lorentz- and CPT-breaking
term $\,K(g_{\mu\nu},k_F)$.

\section{Conclusions}

Quantum effects and, in particular, renormalization, represent an
essential part of the development of the theories with Lorentz- and
CPT-breaking. In the first papers \cite{Kost-loop,ZhLM} the
calculations have been performed by means of Feynman diagrams.
Later on, the functional methods, such as Schwinger-DeWitt
and heat-kernel technique, have been used in \cite{CPTLorentz10}.
In this paper the renormalization has been carried out in curved
space-time and some general features of the renormalization were
established. However, the calculations were not complete, because
only the dimensional symmetry-violating parameters were considered.
In the present paper we go beyond the framework of Ref.
\cite{CPTLorentz10} and derive, for the first time, the
contribution of the dimensionless parameter $k_F^{\mu\nu\al\be}$
in the photon sector to the renormalization of the vacuum.

The performed calculations are new in the sense that we had to
work out the new type of non-minimal operator (\ref{nm}),
different from the standard ones which were considered before
\cite{bavi85}. In these standard cases the non-minimality was
caused by the choice of gauge-fixing parameters. The corresponding
operator can be always studied by integrating over such parameters
starting from the special minimal operator case. In the case of
non-minimal operator (\ref{nm}) the non-minimality is caused by
the presence of external dimensionless {\it function} and this
makes a direct application of the methods of \cite{bavi85}
impossible. The problem has been solved by a trick of inverting
minimal operator and by working in the first order in the
symmetry-violating function $k_F^{\mu\nu\al\be}$. As a result
of this procedure one can start using the functional traces of
\cite{bavi85} and finally arrive at the first-order counterterms.
The obtained expression Eq. (\ref{div}) represents only  a part of
an infinite expansion, according to a general analysis given in
\cite{CPT-ren-2013}. The result also passed a technically
complicated test related to the local conformal invariance.

The derivation of anomaly and anomaly-induced effective action
did not meet serious obstacles, and finally the expression
(\ref{finaction}) was obtained. It turns out that the
dimensionless parameter $k_F^{\mu\nu\al\be}$ makes no
contribution to the dynamics of the conformal factor of the
metric. At the same time, depending on the choice of this
parameter, one can expect a relevant contributions and maybe
even the growth of the tensor modes of metric perturbations
during inflationary epoch. The study of this potentially
interesting problem will require significant efforts, but finally
it can lead to some constraints on the parameter $k_F^{\mu\nu\al\be}$.

\section*{Acknowledgements}
Authors are grateful to Alan Kosteleck\'y for useful
correspondence. This work was partially supported by CAPES,
CNPq, FAPEMIG and ICTP (I.Sh.).

\renewcommand{\baselinestretch}{0.9}

\end{document}